\documentclass[twocolumn,showpacs,amsmath,amssymb,prb]{revtex4-1}
\usepackage{graphicx}
\usepackage{bm}
\usepackage{dcolumn}

\begin{document}
\title{Effect of Curvature on the Electronic Structure and Bound State Formation in Rolled-up Nanotubes}

\author{Carmine  Ortix$^{1,2}$ and Jeroen van den Brink$^{2,3}$}

\affiliation{$^{1}$Institute Lorentz for Theoretical Physics, Leiden University, P.O. Box 9506, 2300 RA Leiden, The Netherlands}
\address{$^{2}$Institute for Theoretical Solid State Physics, IFW Dresden, D01171 Dresden, Germany}
\affiliation{$^{3}$Institute for Molecules and Materials, Radboud Universiteit Nijmegen, 6500 GL Nijmegen, The Netherlands}

\date{\today}

\begin{abstract}
We analyze the electronic properties of a two-dimensional electron gas rolled-up into a nanotube by both numerical and analytical techniques.  The nature and the energy dispersion of the electronic quantum states strongly depend upon the geometric parameters of the nanotube: the typical radius of curvature and the number of windings. The effect of the curvature results in the appearance of atomic-like bound states localized near the points of maximum curvature. For a two-dimensional sheet rolled up into an Archimedean spiral we find  that the number of bound states is equal to the number of  windings of the spiral.
\end{abstract}

\maketitle

\section{Introduction}
\label{sec:intro}
As a consequence of relaxation of elastic stresses,  a thin solid film that is subject to compressive strain, curls up after being partially released from its substrate \cite{pri00,sch01}. This occurs when the strain gradient across the film thickness is sufficiently large to overcome the tendency to form wrinkles, which appear in the opposite limit of small strain gradient\cite{cen09}. The roll-up of a thin solid film into a rolled-up nanotube (RUNT) is particularly exciting since RUNTs have a unique structure \cite{kra06,den06,den09}  that mimic the cylindrical symmetry of a radial crystal. This is reflected in their peculiar magnetic \cite{sha07,fri07,vor07,den09}  and optical \cite{kip06} properties. Moreover, RUNTs are promising candidates for applications in fields ranging from nanofluidics to optics \cite{den04,son07,ber08,smi09}. The experimental progress in manufacturing rolled-up nanostructures triggers the need for a comprehensive theoretical understanding of the quantum carrier dynamics in curved nanomaterials.  

The formal description of the quantum motion of a particle confined to a curved surface was a puzzle for a long time. The problem arises because Dirac quantization in a curved manifold leads to operator ordering ambiguities \cite{dew57}. The situation was cleared up by Da Costa in Ref.~\onlinecite{dac82}. The formal description becomes well-defined when the confinement of the particle on a curved $n$-dimensional manifold is treated as the limiting case of a particle in a $n+1$ dimensional manifold that has a confining force acting in the normal direction of its $n$-dimensional surface.  Because of the lateral confinement, quantum excitation energies in the normal direction become much higher than in the tangential direction. Henceforth, one can safely ignore the particle motion in the direction normal to the surface and on the basis of this deduce an effective, dimensionally reduced Schr\"odinger equation. This procedure is obviously the most rigorous and physically sound one for two-dimensional (2D) curved systems embedded in an ordinary euclidean three-dimensional space. In this case one finds that due to the curvature a scalar potential of purely quantum nature appears in the effective 2D Hamiltonian. Its magnitude is related to the local surface curvature \cite{dac82} so that the quantum mechanics of particles confined to thin curved layers is different from those on a flat plane. Several studies have analyzed the influence of the curvature induced Da Costa scalar potential on the electronic states \cite{can00,aok01,enc03,fuj05,kos05,gra05,jens09} and the electron transport properties \cite{cha04,mar05,cuo09} of a number of different curved systems with complex geometrical shapes. Particularly interesting is the interplay of curvature and electron-electron interaction effects \cite{shi09}. 

Here we concentrate on rolled-up nanostructures, in particular in the form of Archimedean spirals. Although single material structures have been proposed \cite{zan07} and even fabricated \cite{son06},  RUNTs are generally made from bilayer or multilayer thin films of different materials, e.g. GaAs/InGaAs. The two-dimensional electron gas (2DEG) in one the layers is thus confined on a cylindrical surface whose cross section can be fairly approximated by an Archimedean spiral $r = l \phi $ where $r$ and $\phi$ are the cylindrical coordinates in the plane perpendicular to the cylinder axis $z$ and $l$ is related to the radial superlattice constant by $a_{r}=2 \pi l$, see Fig.~\ref{fig:runt}(b). The aim of this work is to investigate the single particle states of a 2DEG in a RUNT. The characteristic Coulomb-like form of the curvature induced scalar potential \cite{ved00} implies the appearance of localized, atomic-like states.  We investigate how their corresponding binding energies are related to the length, curvature, and inner radius of the nanotube and proof that the number of these bound states is equal to the number of windings of the spiral.

This paper is organized as follows: in Sec.~\ref{sec:hamiltonian} we introduce the geometry of the system under study and the correspondent effective Hamiltonian; in
Sec.~\ref{sec:boundstates} the theory is applied to calculate spectra and wavefunctions and we conclude in Sec.~\ref{sec:conclusions}.

\section{Hamiltonian of a 2DEG in a RUNT}
\label{sec:hamiltonian}
\begin{figure}
\includegraphics[width=8cm]{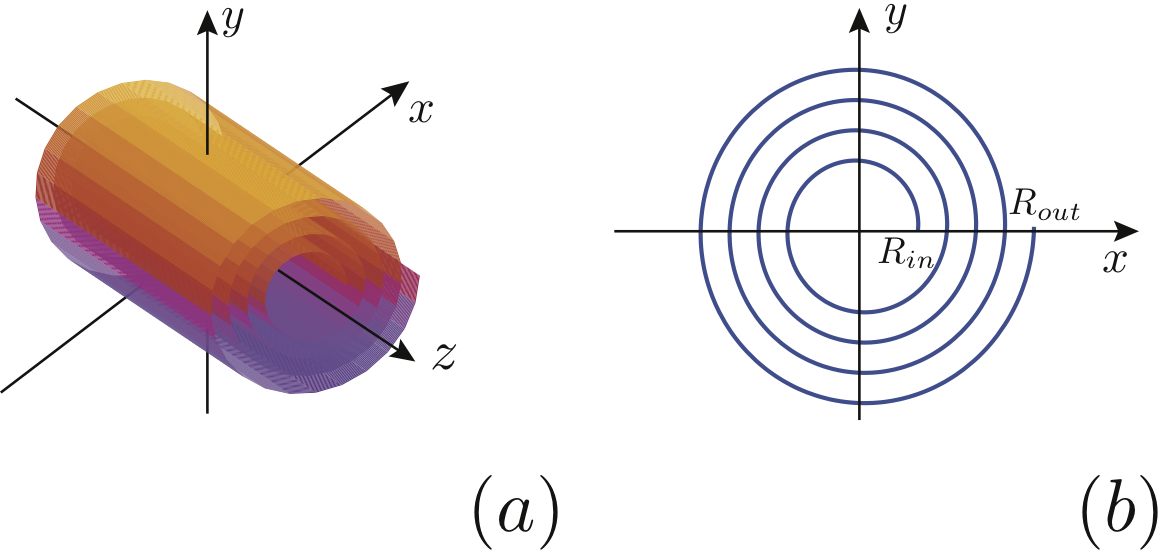}
\caption{(Color online) (a) Sketch of the surface ${\cal S}$ of a RUNT. The Cartesian reference frame we use throughout this paper is indicated. (b) Cross section of the RUNT perpendicular to the cylindrical axis. $R_{in}$, $R_{out}$ correspond to the inner and the outer tube rotation radius respectively.}
\label{fig:runt}
\end{figure}

We first derive the effective Hamiltonian for electrons bound to the surface ${\cal S}$ of a RUNT. As discussed in the previous section, electrons in a RUNT are confined to a cylindrical surface whose cross section can be approximated by an Archimedean spiral [see Fig.~\ref{fig:runt}(b)]. Therefore it is natural to adopt cylindrical coordinates ${\bf r}=\left\{r,\phi,z \right\}$ and parametrize the surface ${\cal S}$ as
\begin{equation}
\left\{ \begin{array}{l}
x=l \,\phi \,\cos{\phi}, \\
y=l \, \phi \, \sin{\phi}, \\
z= z,  
\end{array}
\right.
\label{eq:spiralpareq}
\end{equation}
with $z \in \left(-\infty, \infty \right)$ whereas $\phi \in \left(\phi_{in}, \phi_{out}\right)$. The endpoint of the Archimedean spiral $\phi_{in}$ ($\phi_{out}$) is related to the inner (outer) radius of the RUNT by $R_{in,out}= l  \,\phi_{in,out}$ where $l$ is the typical length scale of the radial superlattice constant $a_r=2 \pi l$ . The maximum radius of the outer tube rotation is instead related to the number of rotations $N_R$ by
$$R_{out}=R_{in}+ 2 \pi\, l\, N_R,$$
where $N_R$ is treated, for convenience, as a continuous variable. From Eq.~(\ref{eq:spiralpareq}), the covariant components of the surface metric tensor are
\begin{equation}
\left\{
\begin{array}{l}
g_{\phi,\phi}=l^2 \, \left(1+\phi^2\right), \\
g_{z,z}=1,   \\
g_{\phi,z}=g_{z,\phi}=0,
\end{array}
\right.
\label{eq:metrictensor}
\end{equation}
whereas the covariant components of the Weingarten curvature tensor \cite{dac82} come out
\begin{equation}
\left\{
\begin{array}{l}
\alpha_{\phi,\phi}= \dfrac{2+\phi^2}{l\, \left(1+\phi^2\right)^{3/2}}, \\
\\
\alpha_{z,z}=\alpha_{\phi,z}=\alpha_{z,\phi}=0.
\end{array}
\right.
\label{eq:weingartentensor}
\end{equation} 
The mean curvature is then given by $M=\alpha_{\phi,\phi} / 2$ whereas the Gaussian curvature is obviously zero.  Following Ref.~\onlinecite{fer08}, the effective 2D Hamiltonian for the tangential motion to the surface ${\cal S}$ becomes:
\begin{eqnarray}
{\cal H}&=&\dfrac{\hbar^2}{2 \, m} \dfrac{1}{\sqrt{g_{\phi,\phi}}} \partial_\phi \left(\dfrac{\partial_\phi}{\sqrt{g_{\phi,\phi}}}\right)-\dfrac{\hbar^2}{8 \, m} \alpha_{\phi,\phi}^2-\dfrac{\hbar^2}{2 m} \partial^2_z 
\label{eq:hamitonian2d}
\end{eqnarray}
where $m$ is the effective mass.
Since the translational invariance along $z$ remains unbroken, the surface wavefunction separates as
$$\Psi(\phi,z)=\psi(\phi) \times \mathrm{e}^{\mathrm{i} k_z \, z},$$
where $k_z$ is the momentum  along the RUNT axis. 
This leads to an effective one-dimensional (1D) Hamiltonian for the $\psi(\phi)$ component of the surface wavefunction 
\begin{equation}
{\cal H}_{1D}= {\cal K}+ {\cal V}_G+ \dfrac{\hbar^2 k_z^2}{2 m},
\label{eq:hamiltonian1d}
\end{equation}
where ${\cal K}$ is the kinetic energy operator for a particle constrained to move along a planar Archimedean spiral waveguide
\begin{equation}
{\cal K}=\dfrac{\hbar^2}{2 \, m \, l^2} \left[-\dfrac{\partial^2_\phi \,}{1+\phi^2} + \dfrac{\phi \, \partial_{\phi} }{\left(1+\phi^2\right)^2} \right],
\label{eq:kinetic}
\end{equation}
whereas
${\cal V}_G$ is the attractive geometric potential induced by the curvature \cite{dac82}
\begin{equation}
{\cal V}_G=-\dfrac{\hbar^2}{8 \,m\, l^2}\, \dfrac{\left(2+\phi^2\right)^2}{\left(1+\phi^2\right)^3}.
\label{eq:geometricpotential}
\end{equation}

One should note that the kinetic energy term Eq.~(\ref{eq:kinetic}) and the geometric potential Eq.~(\ref{eq:geometricpotential}) are different from the expressions derived previously \cite{ved00}  by non-trivial numerical factors. In the following sections, we will find the eigenstates of the Hamiltonian Eq.~(\ref{eq:hamiltonian1d}) by imposing on the $\psi(\phi)$ component of the surface wavefunction Dirichlet boundary conditions at the inner and outer radius of the RUNT and requiring, as usual, square integrability.  

\begin{figure}
\includegraphics[width=0.9\columnwidth]{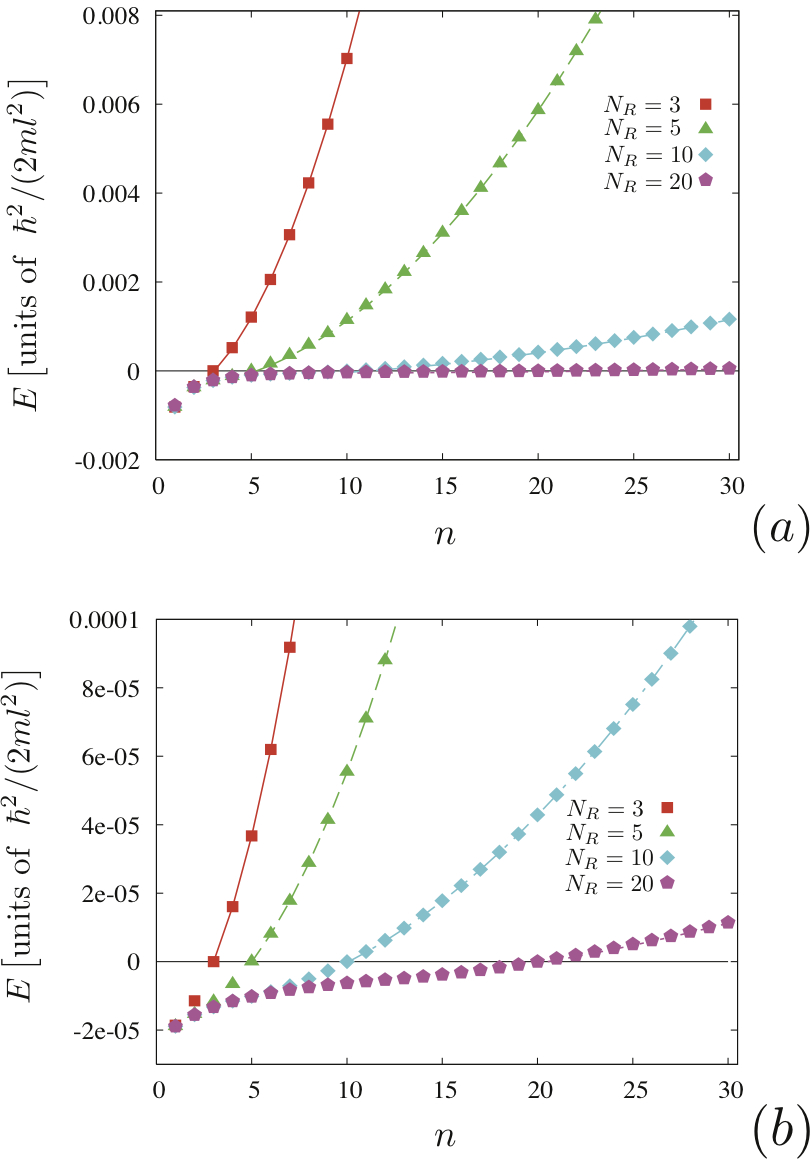}
\caption{(Color online). (a) Electronic spectrum of an Archimedean spiral with inner radius $R_{in}=3 \pi l$ and different numbers of rotations. The points are the numerical results of the exact diagonalization of the Hamiltonian whereas the corresponding continuous lines are the energies obtained considering the effect of the geometric potential in first-order perturbation theory. (b) Same for a RUNT with inner radius $R_{in}=32 \, \pi l$.}
\label{fig:spectrum}
\end{figure}

\section{Curvature induced bound states}
\label{sec:boundstates}
The bandstructure corresponding to the effective 1D Hamiltonian for a 2DEG in a RUNT in Eq.~(\ref{eq:hamiltonian1d})  consists of parabolic subbands 
$$E_{n}(k_z)=E_n^0+\dfrac{\hbar^2 k_z^2}{2 m},$$
with $n$ denoting an integer subband index ($n>0$). Note that the zero of the energy has been fixed at the bottom of the 2DEG conduction band in its planar configuration.
It is then obvious that our problem reduces to the motion of a particle along a planar Archimedean spiral where the subband index $n$ and $E_n^0\,$  respectively label the eigenmodes and the corresponding eigenenergies of the Hamiltonian:
\begin{equation}
{\cal H}^0={\cal K}+{\cal V}_G.
\label{eq:hamiltonian1db0}
\end{equation}  
The exact eigenstates of ${\cal H}^0$ can be found by writing the $\psi(\phi)$ component of the total surface wavefunction as 
\begin{equation}
\psi\left(\phi\right)=\sum_{j=1}^{\infty}\,\, c_j\,\, \chi_{j}\left(\phi\right),
\label{eq:waveexpansion}
\end{equation}
where the $\chi_j\,$'s are the eigenstates of the kinetic energy operator. To proceed further, it is convenient to introduce the arclength of the Archimedean spiral measured from $\phi=0$
\begin{equation}
s(\phi)=\dfrac{l}{2} \left[\phi \left(1+ \phi^2 \right)+ \log{\left(\phi + \sqrt{1+\phi^2}\right)}  \right].
\label{eq:arclength}
\end{equation}
In terms of $s$ the kinetic energy operator takes the compact form ${\cal K}=-\hbar^2 \partial_s^2 / ( 2 m)$ and the corresponding eigenstates can be written as standing waves
\begin{equation}
\chi_j(s)=\sqrt{\dfrac{2}{L}} \sin{\left[\dfrac{\pi \,\,j}{L} \left(s-s_{in}\right)\right]}.
\label{eq:standingwave}
\end{equation}
In the equation above $L$ indicates the total length of the Archimedean spiral whereas $s_{in}$ is the arclength value at the inner radius of the RUNT. By direct diagonalization of the Hamiltonian on the basis of the $\chi_j$'s, we obtain the eigenstates and the corresponding energy spectrum for any value of $R_{in}$ and $N_R$, choosing $l$ as the unit length scale. All reported calculations are obtained introducing a cutoff $j_{max}=100$ in the infinite sum Eq.~(\ref{eq:waveexpansion}), which in all cases is sufficient for convergence.  

As shown in Fig.~\ref{fig:spectrum} the spectrum consists of two distinct regions.
At high energies, the spectrum has a free-particle-like quadratic dependence on $n$ ($E_n^0 \sim n^2$).
In this regime, a good approximation consists in retaining the effect of ${\cal V}_G$ in first-order perturbation theory [continuous lines in Fig.~\ref{fig:spectrum}]. On the contrary, the low energy part of the spectrum  is dominated by the effect of the geometric potential which therefore produces a strong mixing of the free particle states. For integer number of rotations, there is a critical mode that separates the two extreme spectral structures corresponding to $n \equiv N_{R}$ independent of the inner radius of the RUNT. This critical state corresponds to a zero energy state where the geometric potential energy balance the kinetic energy. 

\begin{figure}
\includegraphics[width=0.9\columnwidth]{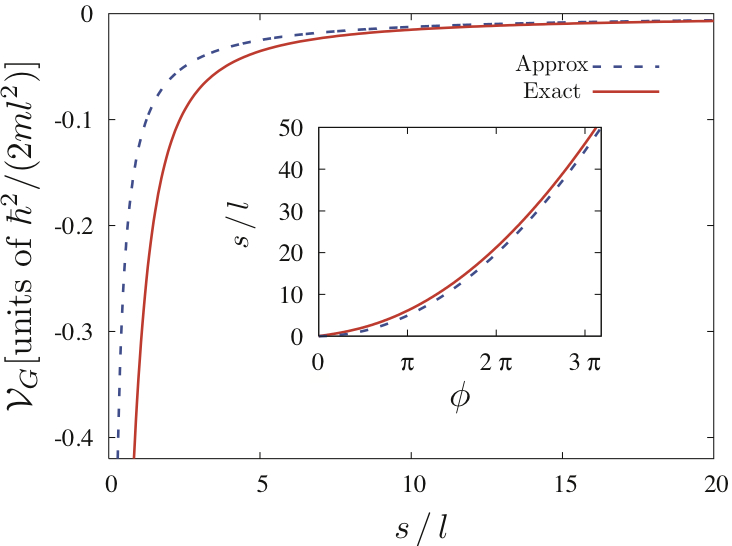}
\caption{(Color online). Comparison of the exact (full line) and the asymptotic  (dashed line) form of the geometric potential measured in units of $\hbar^2 / (2 m l^2)$  as a function of $s / l $. Inset: same for the behavior of the arclength   as a function of the azimuthal angle $\phi$ of the spiral.}
\label{fig:geometricpotential}
\end{figure}

\begin{figure}
\includegraphics[width=0.9\columnwidth]{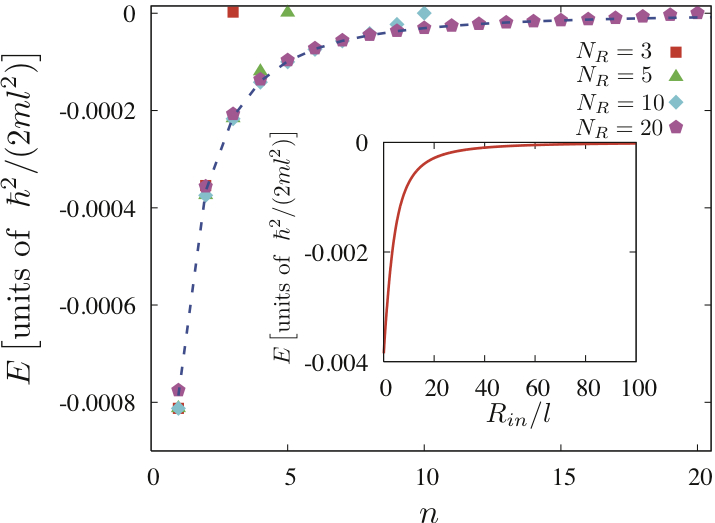}
\caption{(Color online). Behavior of the lowest eigenenergies $E_n^0$ as a function of the quantum number $n$ for a  RUNT with $R_{in}= 3 \pi l$ and different number of rotations $N_{R}$. The points are the numerical results whereas the continuous line corresponds to the analytical behavior of the spectrum of the one-dimensional hydrogen Hamiltonian. The inset shows the behavior of the ground state binding energy as a function of the inner radius of the RUNT.}
\label{fig:hydlevels}
\end{figure}

Next we show that the appearance of these two distinct spectral structures  emerges as a natural consequence of the competition between the confinement due to the Dirichlet boundary conditions at the inner and outer radius of the RUNT and the effect of the geometric potential. 
In order to make a qualitative analysis of the spectrum, it is convenient to consider the asymptotic form of the geometric potential ${\cal V}_G \sim - \hbar^2 / (8 \, m \, l^2 \, \phi^2)$.  
Apart from a logarithmic correction, the arclength of the spiral Eq.~(\ref{eq:arclength}) grows quadratically with $\phi$ [see inset of Fig.~\ref{fig:geometricpotential}]. Then, it turns out that the asymptotic form of the geometric potential in terms of $s$ is a Coulomb one \cite{ved00} [see Fig.~\ref{fig:geometricpotential}]. It is then clear that the geometric potential corresponds to an attraction towards the point of maximum curvature $R_{in}$ and leads to the appearance of bound states. 
The asymptotic form of the Hamiltonian Eq.~(\ref{eq:hamiltonian1db0}) reads
\begin{equation}
{\widetilde {\cal H}}^0= -\dfrac{\hbar^2}{2 m} \partial^2_{s}-\dfrac{\hbar^2}{16 m  l} \,\,\dfrac{1}{s}.
\label{eq:hamiltonianapprox}
\end{equation}
By restricting to the half-space $s \geq 0$, Eq.~(\ref{eq:hamiltonianapprox}) is the Hamiltonian of a 1D hydrogen atom with a ``quantum charge''  $e_q=\hbar / ( 4 \sqrt{m\, l})$. 
The eigenstates and the corresponding eigenenergies  are thus well known. 
However, in the present situation we have to meet the Dirichlet boundary condition at $s_{in}$ and $s_{in}+L$. The effect of these boundary conditions can be captured in a two step process.
 First, the boundary condition at $s_{in}$ is met by the infinite set of localized atomic-like states that, apart from a normalization constant, read
\begin{equation}
\psi_n (s)= \,\, s \, \mathrm{e}^{-s/(\eta_n a_0)} \,\,\, U\left[1-\eta_n\, ,\, 2 ,\,\dfrac{2 \, s}{\eta_n a_0} \right],
\label{eq:hydfunctions}  
\end{equation}
where $U$ is the confluent  hypergeometric function of the second kind and we defined the ``Bohr radius'' $a_0=\hbar^2 / (m e_q^2)=\, 16 \, l$. Finally, the parameters $\eta_n$, which depend on $s_{in}$, determine the binding energies
\begin{equation}
E_n^0=-\dfrac{\hbar^2}{2 \, m \, a_0^2 \,\eta_n^2}.
\label{eq:hydlevels}
\end{equation}
Obviously, for $s_{in}=0$ the energy spectrum reduces to the usual Rydberg series ($\eta_n \equiv n$). By increasing $s_{in}$, the $\eta_{n}$'s grow linearly with $s_{in}$ meaning that the binding energies are inversely related to  the inner radius of the RUNT [see inset of Fig.~\ref{fig:hydlevels}]. 

Next, we introduce the Dirichlet boundary condition at the outer radius of the RUNT. The atomic-like states do not meet this boundary condition since they do not vanish exactly at $s_{out}=s_{in}+L$. However, if $s_{out}$ resides in their exponential tail, the effect of the latter boundary condition can be neglected. This will obviously occur for the lowest energy states for which $s_{out}$  is much larger than the average arclength $\langle s \rangle$. Their corresponding binding energies will be then accurately predicted by Eq.~(\ref{eq:hydlevels}) as shown in Fig.~\ref{fig:hydlevels}. This is not verified for large $n$ since the atomic-like states are localized over a region much larger than the total length of the spiral. The confinement due to the Dirichlet boundary conditions will dominate in the latter case and hence we expect the  exact eigenfunctions to be similar to the standing waves of Eq.~(\ref{eq:standingwave}). 

\begin{figure}
\centerline{\includegraphics[width=.9\columnwidth]{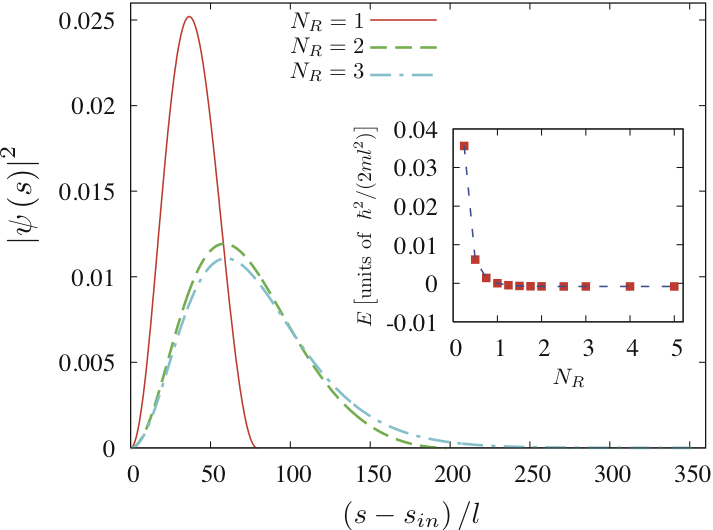}}
\caption{(Color online). Ground state density probability as a function of $s-s_{in}$ for a RUNT with $R_{in}=3 \pi l$ performing up to three rotations. 
The total length of the spiral corresponds to $L \sim 80 l$ for $N_{R}=1$, $L \sim 200 l$ for $N_{R}=2$ and $L \sim 350 l$ for $N_{R}=3$. 
The inset shows the behavior of the corresponding eigenvalue as function of the number of rotations $N_R$. For $N_{R}>1$ the energy of the ground state saturates at a finite value.}
\label{fig:wavegroundstate}
\end{figure}

By increasing the number of rotations $N_R$ or equivalently the total length of the spiral $L$, one then finds a continuous evolution from free particle states where the eigenfunction is localized over the entire length $L$, to
atomic-like states where the localization region is of the order of $\langle s \rangle$  [see Fig.~\ref{fig:wavegroundstate}]. Accordingly, as shown in the inset of Fig.~\ref{fig:wavegroundstate},  the eigenvalue scale with $1 / L^2$ in the free particle region saturating at the finite negative value given by the binding energies Eq.~(\ref{eq:hydlevels}). 

Now we can determine the appearance of zero energy eigenstates. The asymptotic Hamiltonian Eq.~(\ref{eq:hamiltonianapprox}) admits a zero energy eigenstate which has, apart from a normalization constant,  the following general form
\begin{equation}
\psi_{0}(s)= \, \sqrt{\dfrac{s}{l}}\, J_{1} \left(\sqrt{\dfrac{s}{2 l}}\right)+ C \, \sqrt{\dfrac{s}{l}} \, Y_{1} \left(\sqrt{\dfrac{s}{2 l}}\right).
\label{eq:zeroenergyeigenstate}
\end{equation}
In the equation above, $J$ and $Y$ indicate respectively the Bessel functions of the first and second kind whereas $C$ is an arbitrary constant that can be fixed by requiring Eq.~(\ref{eq:zeroenergyeigenstate}) to meet the Dirichlet boundary condition at the inner radius of the RUNT. Here it is convenient to write the eigenstate Eq.~(\ref{eq:zeroenergyeigenstate}) in terms of the azimuthal angle of the Archimedean spiral. Since in the $\phi>>1$ regime $s \sim l \,\phi^2 / 2$ [see Eq.~(\ref{eq:arclength})],  we find
\begin{equation}
\psi_{0}(\phi) \sim \, \sqrt{\phi} \left[ \cos{\left(\dfrac{\phi}{2}- \dfrac{3 \pi}{4}\right)}+C \,  \sin{\left(\dfrac{\phi}{2}- \dfrac{3 \pi}{4}\right)} \right]
\label{eq:zeroenergyeigenstateapprox}
\end{equation}
where we got rid of the Bessel functions appearing in Eq.~(\ref{eq:zeroenergyeigenstate}) by taking advantage of their asymptotic expansion for large $\phi$ values \cite{abr64}. From Eq.~(\ref{eq:zeroenergyeigenstateapprox}) it is immediately clear that the zero energy eigenstate meets the second Dirichlet boundary condition at the outer radius of the RUNT only for $\phi_{out}=\phi_{in}+ 2 \pi k$ with $k$ integer and hence for an integer number of rotations independent  of the inner radius of the RUNT. Notice that the zero energy state will have $N_{R}-1$ nodes and thus will represent the $N_R$-th lowest energy state as indeed numerically found.  Thus the number of curvature-induced bound states is equal to the number of windings of the Archimedean spiral.

\section{Conclusions}
\label{sec:conclusions}
In conclusion, we have investigated theoretically the single particle states in a rolled-up nanotube and have found that the effect of the curvature results in the appearance of atomic-like localized states. Interestingly the number of the bound states corresponds to the rotation number of the nanotube. We have also determined how the binding energies depend on the other relevant geometric parameters, namely, the radial superlattice constant and the typical radius of the nanotube. 

\section*{Acknowledgments}
The authors are pleased to thank V. Fomin, S. Kiravittaya and O.G. Schmidt for fruitful discussions. This work was supported by the Dutch Science Foundation (FOM).

\end{document}